\documentclass[12pt]{article}
\usepackage{graphicx}
\usepackage{graphics}
\usepackage{epsfig}
\usepackage[utf8x]{inputenc}
\usepackage{subfigure}
\usepackage{bm}
\usepackage{newlfont}
\begin{document}

\begin{flushright}
RECAPP-HRI-2012-001
\end{flushright}

\begin{center}
{\Large\bf Constraints on invisible Higgs decay in MSSM
in the light of $h^0 \longrightarrow\gamma\gamma$ rates from the LHC}\\[20mm]

Nishita Desai \footnote{E-mail:
nishita@hri.res.in}, Biswarup Mukhopadhyaya\footnote{E-mail:
biswarup@hri.res.in} and Saurabh Niyogi \footnote{E-mail:
sourabh@hri.res.in}\\

{\em Regional Centre for Accelerator-based Particle Physics,\\
Harish-Chandra Research Institute,\\
Chhatnag Road, Jhusi, Allahabad - 211 019, INDIA}

\end{center}

\begin{abstract}
We examine the parameter space of the purely phenomenological minimal
supersymmetric standard model (MSSM), without assuming any supersymmetry
 breaking scheme.  We find that a large region of the parameter space can indeed
yield the lightest neutral Higgs mass around 125 GeV, as suggested by
the recent ATLAS data, and also lead to event rates around, or
slightly higher than, the standard model expectation in the two-photon
and four-lepton channels. Using a lightest neutralino that is
considerably lighter than the Higgs, we find that the `invisible' decay
of the Higgs into a pair of neutralinos upto about 10\% can be consistent 
with the current data from the Large Hadron Collider (LHC).

\end{abstract}

\newpage

\section{Introduction}
The most recent results from the Large Hadron Collider (LHC) keep the
hopes for an imminent discovery of the Higgs boson healthily
alive. The search results based on an integrated luminosity of 4.9
$fb^{-1}$ are already available. While the mass range approximately
between 115 and 127 GeV cannot be ruled out for a standard model (SM)
Higgs, the ATLAS collaboration reports hints of an excess around 125
GeV, in both the $\gamma\gamma$ and $ZZ^*$ channels
\cite{ATLAS1,ATLAS2,ATLAS3,ATLAS4,ATLAS5}.  It may be premature to
read too much into this suggested `peak', but it has quite
understandably, raised hopes which are reflected in a large number of
theoretical papers written since the announcement of the results
\cite{hall,Baer,matchev,Hein,Arbey1,Arbey2,Draper,moroi,Carena,
Ellwanger,akula,raidal,Espinosa:2012ir}.

As of now, the observed signal in, for example, the $\gamma\gamma$
channel is more or less consistent with the standard model
expectation, when the Higgs mass is about 125 GeV. At the same time,
it is also important to know what the data implies for physics beyond
the standard model. The very existence of a low-lying electroweak
symmetry breaking sector raises the naturalness and triviality
problems, and calls for new physics explanations for the stability of
the electroweak scale. The scenario that has drawn maximum attention
in this context is supersymmetry (SUSY) \cite{nilles,haber,martin}
which in its minimal phenomenological form (MSSM) is still awaiting a
complete probe at the LHC. The fact that the lightest neutral Higgs
boson can at most be of mass about 135 GeV makes the currently allowed
Higgs mass range a subject of even closer attention from the viewpoint
of SUSY. The recent studies
\cite{hall,Baer,matchev,Hein,Arbey1,Arbey2,Carena,akula,raidal,Bottino} 
consider the implications of the Higgs result in various SUSY scenarios. The
present study is another step in this direction, where we have relaxed
the requirement of any simplifying SUSY breaking scheme.

The rate in the $\gamma\gamma$ channel, having a small branching
ratio, tends to get suppressed when other channels receive any
substantial boost. It has been, therefore, argued that the visibility
of the Higgs in this channel can be reduced due to enhancement of the
$h{\bar b}b$ coupling (where h is the lightest neutral CP-even
scalar). This reduction can be compensated, for example, by enhancing
the stop mixing angle or by reducing the lighter stau mass
\cite{Carena}.  The detailed study of MSSM parameter space that
survives the Higgs mass window is presented in \cite{Arbey2}.

If the lightest neutralino ($\chi^0_1$) is stable, it can serve as a
candidate for cold dark matter. At the same time, if its mass is less
than half the Higgs mass, the decay $h \longrightarrow \chi_1^0
\chi_1^0$ can have a large branching fraction, which in turn may also
suppress the $h \longrightarrow \gamma\gamma$ branching fraction.  We
can, therefore, use this information to restrict the MSSM parameter
space based on recent hints of Higgs signatures.  Though, the
importance of the $\chi^0_1 \chi^0_1$ has been pointed out in an
earlier study \cite{vasquez1}, we would like to emphasize its special
role when the lightest SUSY Higgs lies in the region 123-127 GeV in
the light of the results from the ATLAS collaboration. We would also
like to investigate whether it is possible to have SM-like rates in
the $\gamma\gamma$ channel even with significant invisible branching
fraction.  In order to have substantial invisible decay, with the
Higgs in the range 123 - 127 GeV, the neutralino has to be less than
$\simeq$ 50 - 60 GeV. Such values are consistent with the LEP data and
all other constraints in the purely phenomenological MSSM.

The most stringent bound on the mass of dark matter particle available
till date is from the direct detection experiment XENON100, which
imposes bounds on the scattering cross-section versus WIMP mass in the
entire region upto 1 TeV \cite{xenon1,xenon2}.  Recent analysis by
CRESST \cite{cresst} in the low mass region seems to corroborate
DAMA/LIBRA \cite{dama} and CoGeNT \cite{cogent} results in favouring
light dark matter.  Given the stringent bounds in this region from
XENON100, there is an obvious tension in the result from the different
DM detection experiments.  We therefore consider the region with
$m_{\chi^0_1} \sim$ 10-50 GeV without imposing any dark matter related
constraint. For a study based exclusively on DM constraint, we refer
the reader to \cite{mambrini}. Very recently, after the publication 
of ATLAS data, some study related to SUSY dark matter has been performed
in \cite{vasquez2,Ellis:2012aa}.

The questions we propose to answer in this study are as
follows. First, do any regions of the MSSM parameter space (with
$2m_{\chi^0_1}<m_h$), satisfy the Higgs mass constraint and at the
same time reproduce the measured signal strengths in the
$\gamma\gamma$ and $ZZ^*$ channels? Second, what regions of MSSM
parameter space reproduce SM-like signal strengths in the
$\gamma\gamma$ channel. Finally, whether it is still possible to have
a significant invisible Higgs decay when the Higgs mass is within
123-127 GeV and $\gamma\gamma$ signal strength is unsuppressed.

Since the phenomenology pertaining to the Higgs sector depends little
on the gluino and the first two family sfermion masses, we have
assigned large masses to these states, falling back on decoupling as
the means of ensuring the suppression of flavour-changing neutral
currents.  In the same spirit, we have chosen $\mu$ to be completely
free, and not imposed any condition as such radiative electroweak
symmetry breaking to restrict it.

It should be noted that, although our analysis is inspired by dark
matter considerations, relying on a light neutralino LSP in the MSSM,
it also constrains the parameter space of R-parity violating SUSY as
well (where R-parity is defined by $R = (-)^{3B + 2S + L}$).  While
the violation of R-parity can render the lightest neutralino unstable
and destroy its candidature for dark matter constituent, a light
enough neutralino can still eat heavily into the decay width of the
lightest Higgs $h$ and thereby reduce its decay rate in the
$\gamma\gamma$ channel.  Therefore, if the data on this channel from
the LHC continue to maintain the current trend, this channel may
successfully probe an R-parity violating SUSY scenario as well.

The strategy adopted for our analysis is outlined in section 2.
Section 3 contains the constraints we are able to derive on the MSSM
parameter space, with the lightest neutral Higgs capable of decaying
into a pair the lightest neutralino. We summarise and conclude in
section 4.

\section{Strategy for analysis}

As has been already stated, we primarily focus on the $\gamma\gamma$
data published recently.  The rates in loop-suppressed channels such
as $\gamma\gamma$ get substantially affected if the tree-level
couplings of the Higgs are altered due to physics beyond the standard
model.  However, in certain regions of the model parameter spaces, the
interplay between the production cross-section and branching ratio can
reach an overall rate close to, or, even greater than the SM prediction.
Our aim is to first identify such regions in the MSSM parameter space,
and then to check if a substantial invisible branching ratio of the
Higgs in these regions is still possible.

For calculating the Higgs masses and mixing in the MSSM, we have used
the code {\tt FeynHiggs} v $2.8.6$ \cite{degrassi,Hein2,Hein3,Hein4}
which has full two-loop results \cite{Hein5}.  First, we scan the
MSSM parameter space and select only those points for which the mass
of the lightest neutral scalar Higgs ($m_h$) in the range 123 - 127
GeV.  For any given value of $\tan\beta$, the ratio of the vacuum
expectation values (vev) of the two Higgs doublets, the neutral
pseudoscalar Higgs mass ($m_A$) is varied appropriately to achieve
this, and a scan over the permissible values of $m_A$ is performed in
our analysis.  It should be kept in mind that $m_h$ in the
aforementioned range is crucially dependent on radiative corrections
to the potential, largely derived by the top quark and its
superpartner.  Two-loop corrections to the scalar potential are
included for this purpose.

The gluino mass has been fixed at 2 TeV.  Squarks of the first two
families, too are held fixed at a high mass of 2 TeV, just for
simplicity, since they have little bearing on the phenomenology of the
Higgs at the LHC. The sleptons are all held fixed at 800 GeV, and the
diagonal elements of the stop and sbottom mass matrices, at 1 TeV. The
quantities that have been adjusted to obtain the lightest neutral
Higgs mass in the desired band are $\tan\beta$ and $A_t$, the
trilinear soft SUSY breaking parameter in the stop sector. The
$A$-parameters in all other sfermion sectors are set to zero.

With the above choice of parameters, we next watch the role of the
invisible channel $h \longrightarrow \chi_1^0 \chi_1^0$ .  In
particular, we ask the question: is it possible to be consistent with
the current LHC data but still be consistent with a substantial
invisible branching fraction for $h$?  For this, first of all, one
requires $m_h > 2 m_{\chi_1^0}$. Secondly, the composition of
$\chi_1^0$ is important in determining the invisible branching
ratio. Though $\mu$ on the lower side is mostly helpful for this
purpose, we vary it over the entire range 100 GeV - 1 TeV, taking care
at the same time to confine $m_h$ to the pre-decided range.  We hold
the $M_1$, the U(1) gaugino mass at the representative value of 50 GeV
and vary $M_2$, the SU(2) gaugino mass across 100 - 500 GeV. We have
checked that the conclusions do not differ significantly for values of
$M_1$ in the range 10-50 GeV.  Two values of $\tan\beta$, namely, 10
and 40, are used.  We use low-scale values of all parameters in our
scan.  It should be noted that the branching ratio for
$h\longrightarrow b{\bar b}$ can be as low as 7 - 10\% in some regions
of the parameter space.  This is done most effectively through
appropriate values of the stop mixing parameter $A_t$. Therefore, the
effect attempted in reference \cite{Carena}, namely, reducing the
$b{\bar b}$ decay width as much as possible, is included in our
analysis.

The results presented in the conference note corresponding to the
ATLAS analysis \cite{ATLAS5} for 4.9 fb$^{-1}$ data give the signal
strengths in individual channels where an excess over the background
has been observed.  Since the production and decay kinematics for SUSY
Higgs are not different from a SM Higgs, we assume that the detector
efficiency remain the same and therefore the signal in case of a SUSY
Higgs can simply be obtained by taking the ratio of the production
cross section and the branching ratios in the channel under
consideration.

The dominant Higgs production channel is $gg \longrightarrow h$.
Substantial increase to $b\bar b \longrightarrow h$ can also be
observed in certain regions of parameter space.  The next highest
contribution to the cross section is in the vector-boson fusion
channel which contributes to less than 8\% of the total.  We therefore
use the NLO calculation of $gg \longrightarrow h$ and $b\bar b
\longrightarrow h$ available in FeynHiggs to calculate the ratio
$R_{\gamma \gamma}$ defined by :
\newline

\begin{equation}
R_{\gamma\gamma} = \frac{[\sigma(pp \rightarrow h)\times BR(h \longrightarrow \gamma\gamma)]_{MSSM}}{[\sigma(pp \rightarrow h)\times BR(h \longrightarrow \gamma\gamma)]_{SM}}
\end{equation}

\noindent
\newline
A similar quantity, named $R_{ZZ^*}$, is defined for Higgs decay into
the four-lepton channel via a real and a virtual Z.  Since these two
are the primary channel in which a signal has been observed, We use
these ratios in the next section to determine the favoured MSSM
parameter space and the correlation to the invisible Higgs branching
ratio.

\section{Results}

\begin{figure}[t]
\centering
\includegraphics[width=80mm]{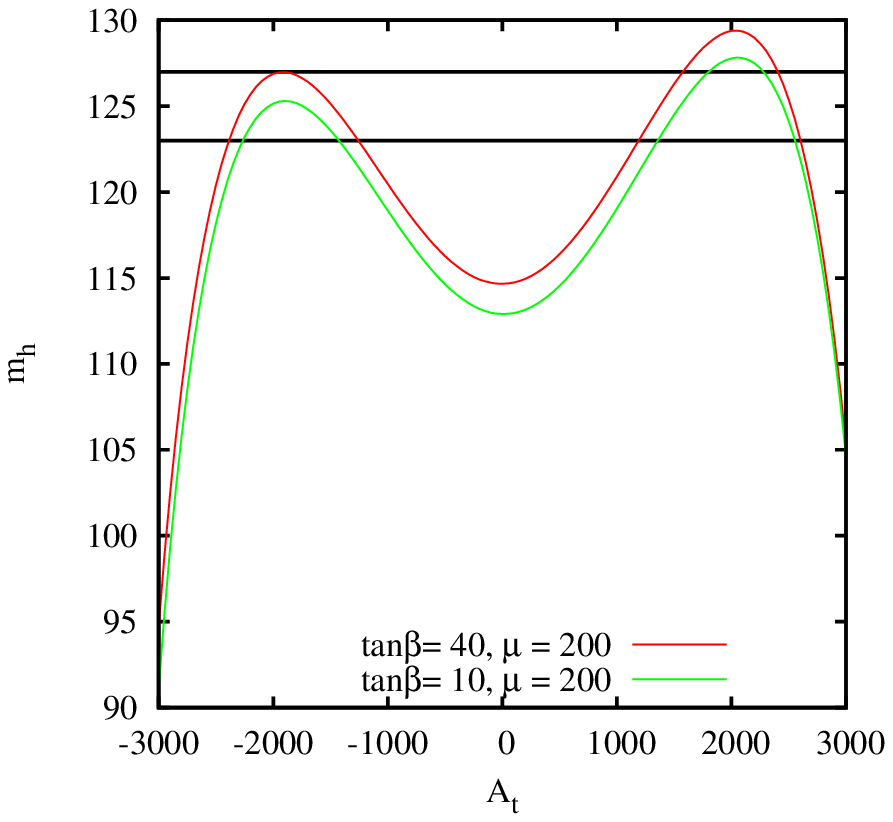}
\includegraphics[width=80mm]{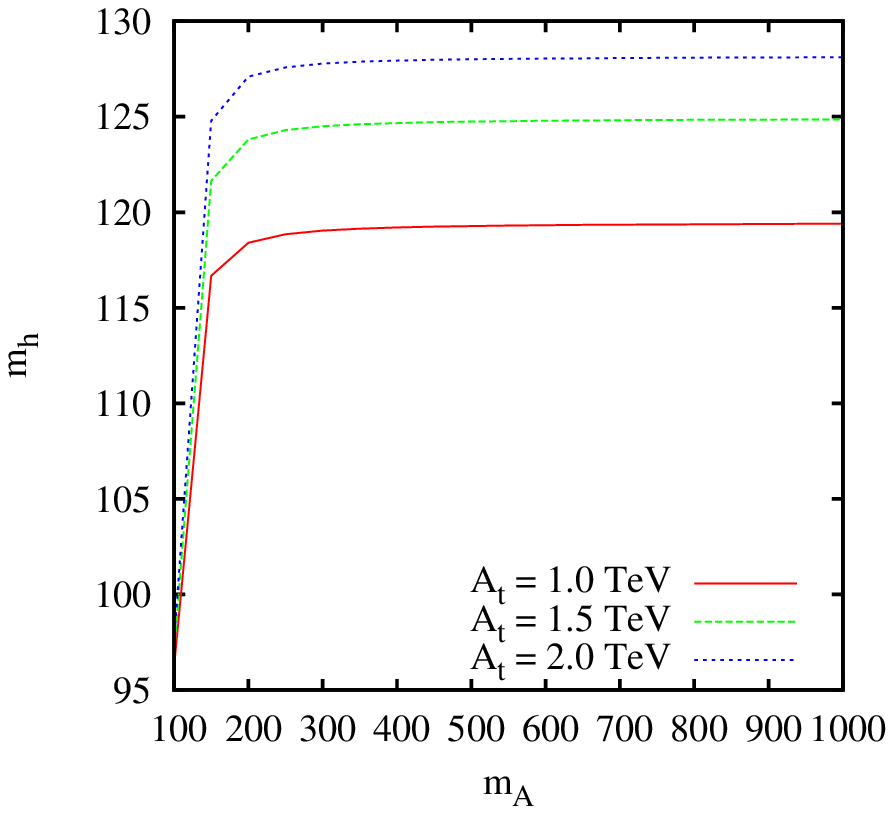}
\includegraphics[width=80mm]{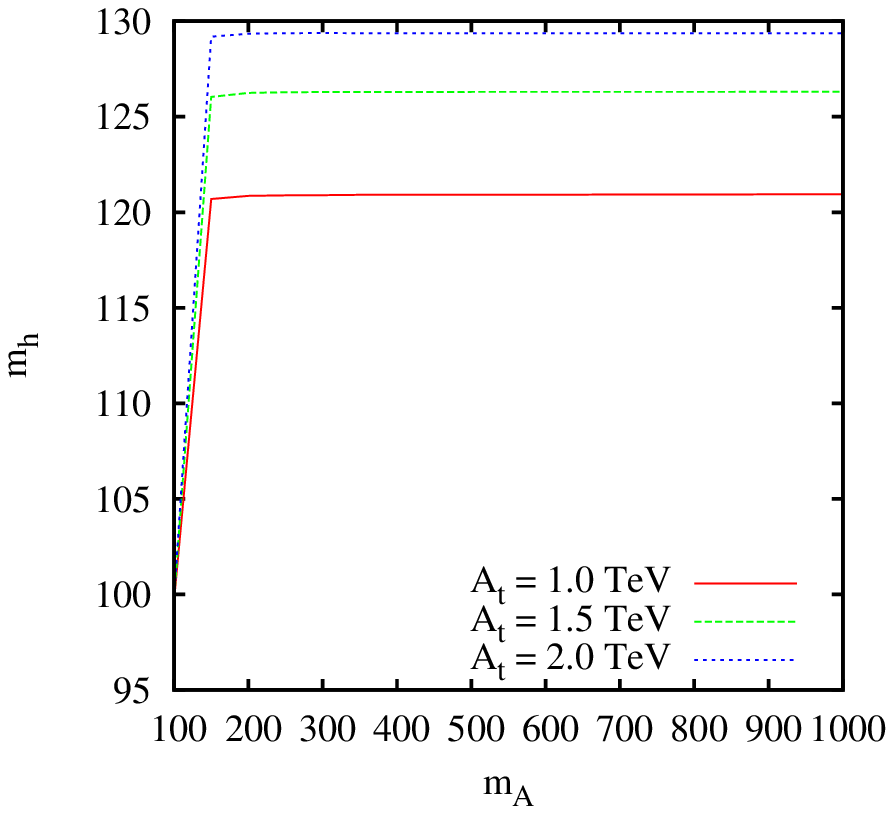}
\caption{ {\footnotesize Dependence of Higgs mass $m_h$ on $m_A$, $A_t$ and $\tan
  \beta$ for $\mu$ = 200 GeV. The first panel (top-left) shows the
  dependence of $m_h$ on $A_t$ for different values of $\tan\beta$
  with. The horizontal band in fig 1(a) corresponds to the `favoured'
  region seen by ATLAS around $m_h$ = 125 GeV. 1(b), (c) are $m_A$ vs
  $m_h$ curve for various $A_t$ and $\tan\beta$ = 10 and 40
  respectively.} \label{fig1} }
\end{figure}

The first objective is to keep $m_h$, the MSSM lightest neutral Higgs
mass, in the neighbourhood of 125 GeV.  The Higgs mass is affected
strongly by the mass of the CP-odd neutral Higgs $m_A$ and the
trilinear stop coupling $A_t$ and the dependence can be seen from
Figure~\ref{fig1}.  The first panel shows the dependence on $A_t$ for
different values of $\tan \beta$.  Restricting $m_h$ to lie in the
window 123-127 GeV restricts the values of $A_t$ to lie within
$\pm(1.2-2.5)$ TeV.

The dependence on $m_A$ can be seen from the second ($\tan \beta =
10$) and third ($\tan \beta = 40$) panels of Figure~\ref{fig1}.
Increasing $\tan \beta$ reduces the $m_A$ required to reach the
maximum Higgs mass value.  We also find that the variation of $m_A$
upward of 300 GeV has practically no effect on $m_h$ for the entire
allowed range of $\tan \beta$.  Although, it does affect the event
rates in various channels driven by the production of $h$.  Therefore,
for further study, we look into three different regions; first with
$m_A = 300,~\mathrm{and~} 1000$~GeV which correspond to the beginning
of the region with maximum Higgs mass and the decoupling limit
respectively, and the low-$m_A$ region with $m_A < 300$~GeV.  We use a
value of $\mu =200$ GeV for illustration in Figure~\ref{fig1}, but,
the qualitative nature of the curves do not change for higher values
of $\mu$.

\begin{figure}[t]
\centering
\includegraphics[width=80mm]{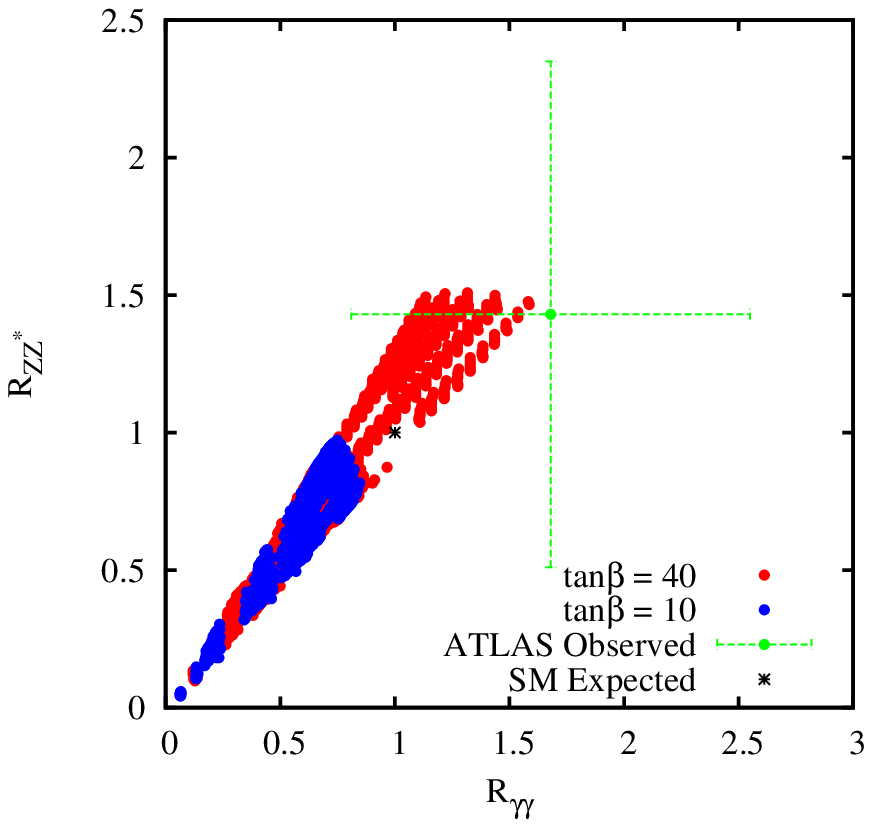}
\includegraphics[width=80mm]{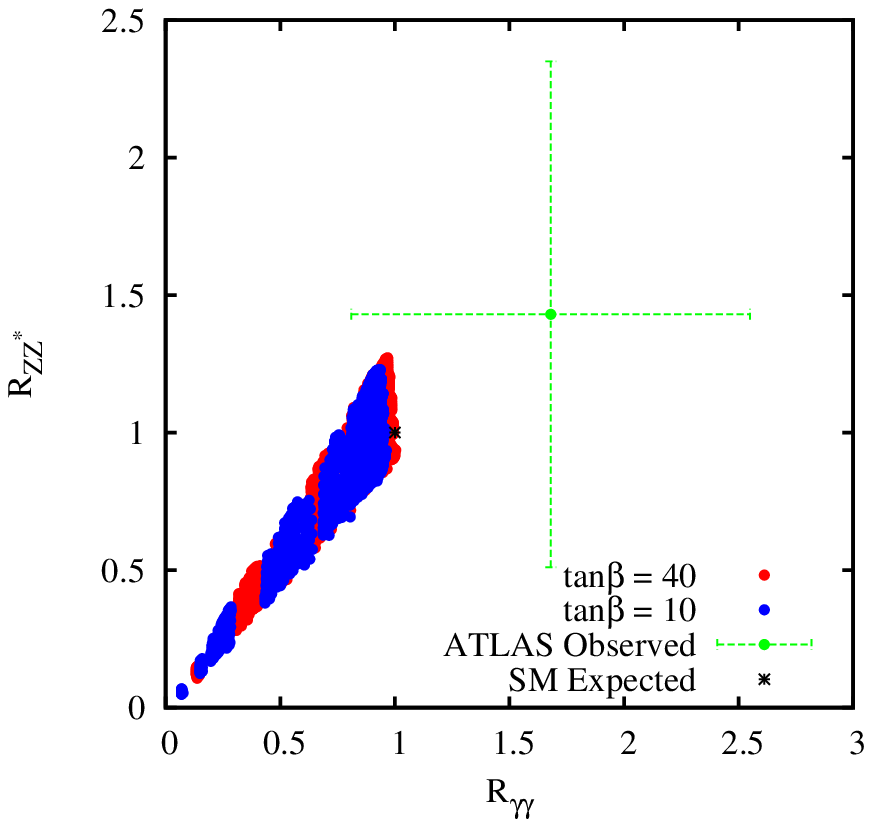}
\includegraphics[width=80mm]{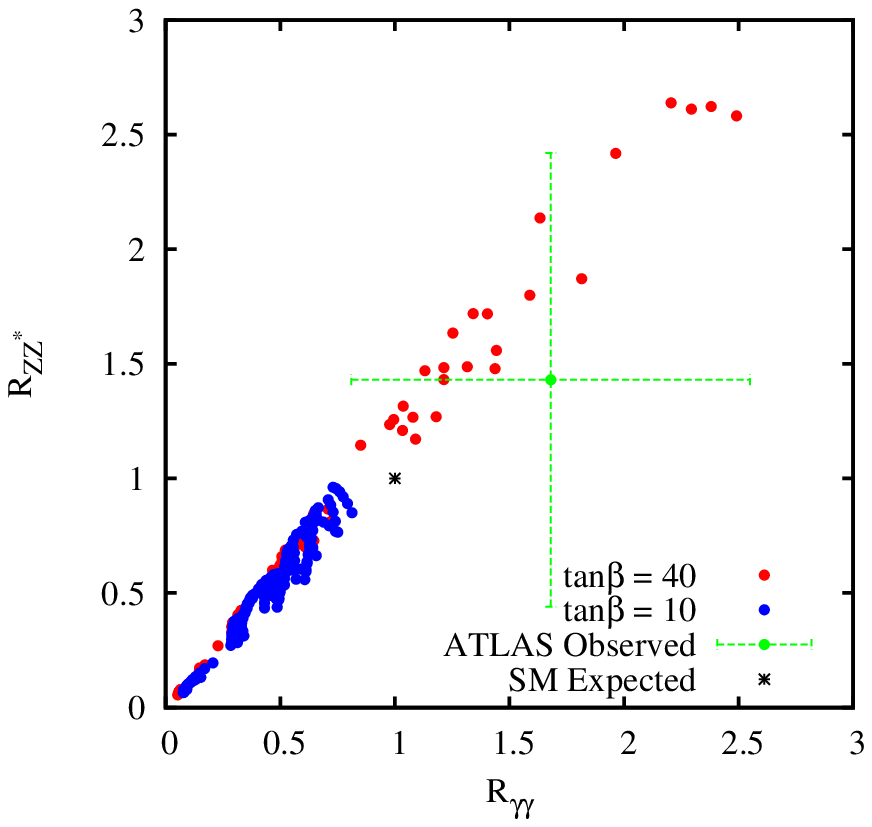}

\caption{ { \footnotesize The values of ratios $R_{\gamma\gamma}$ and $R_{ZZ^*}$ for
  the points in our scan that lie within 123~$<m_h<$~127 GeV.  We use
  $M_1=50$ GeV for illustration and scan over other parameters
  (details in the text).  The top left panel corresponds to $m_A=300$
  GeV, the top-right to $m_A=1$ TeV and the bottom panel the the
  region with $m_A< 300$ GeV.  The ATLAS point corresponds to the
  reported signal strength at the best-fit point with the 1-sigma
  error bands.  The SM point is shown as a black star}. \label{fig2}}
\end{figure}

We now turn our attention to the question of what region is favoured
by the current data.  Fig.~\ref{fig2} shows the scatter of scanned
points in $R_{\gamma \gamma}$ and $R_{ZZ^*}$.  The ATLAS observed
point corresponds to the observed signal strength in the $\gamma
\gamma$ and $ZZ^*$ channels.  The error bars correspond to the 1-sigma
errors reported by the experiment.  We allow only those points that
satisfy the Higgs mass window of 123--127 GeV. The two top panels
correspond to $m_A=300$ GeV and $m_A=1$ TeV respectively.  These two
correspond to the region where the Higgs mass has reached it's maximum
value for a given $A_t$.  We can clearly see that even though the
Higgs mass does not change appreciably in the region (see
Fig.~\ref{fig1}), the branching ratios change considerably.  A lower
value of $m_A$ corresponds to larger values of both $R_{\gamma
  \gamma}$ as well as $R_{ZZ^*}$.  We also see that these two
variables are highly correlated and therefore data in the other Higgs
decay channels like $b\bar b$ and $\tau \bar \tau$ will be crucial to
realistically rule out any MSSM points.  The final panel in
Fig.~\ref{fig2} corresponds to the region $m_A<300$ GeV.  This is the
only region that is capable of reproducing rates close to the observed
ATLAS data point.

In each case, we have used $M_1 =$ 50 GeV, and scanned $M_2$ and $\mu$
(the Higgsino mass parameter) over the range 100-500 GeV, and 100-1000
GeV, respectively. Both $\tan\beta =$ 10 and $\tan\beta =$ 40 have
been included in the scans.  We find that for $\tan \beta = $10, it is
not possible to achieve values of $R_{\gamma\gamma}$ or $R_{ZZ}$
greater than the SM value of unity irrespective of $m_A$.  Therefore,
the currently measured data favours a larger value of $\tan \beta$ in
MSSM.  It should also be mentioned that the scatter plots in the
$R_{\gamma\gamma} - R_{ZZ}$ space are not appreciably different when
one reduces the mass of the lightest neutralino to 10 GeV. Thus our
conclusions are unaltered even for a relatively light dark matter
candidate.

\begin{figure}[t]
\centering
\includegraphics[width=80mm]{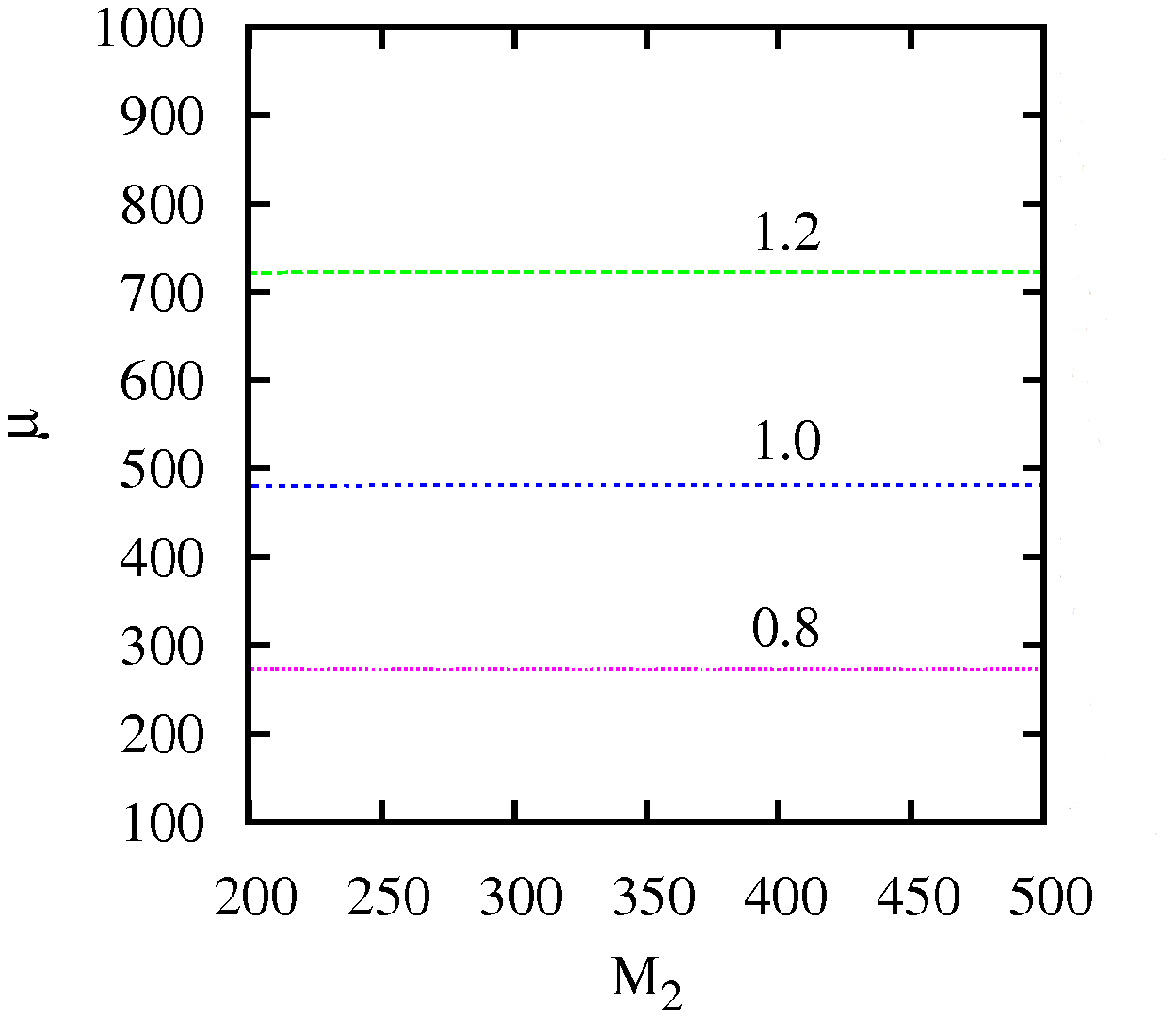}
\includegraphics[width=80mm]{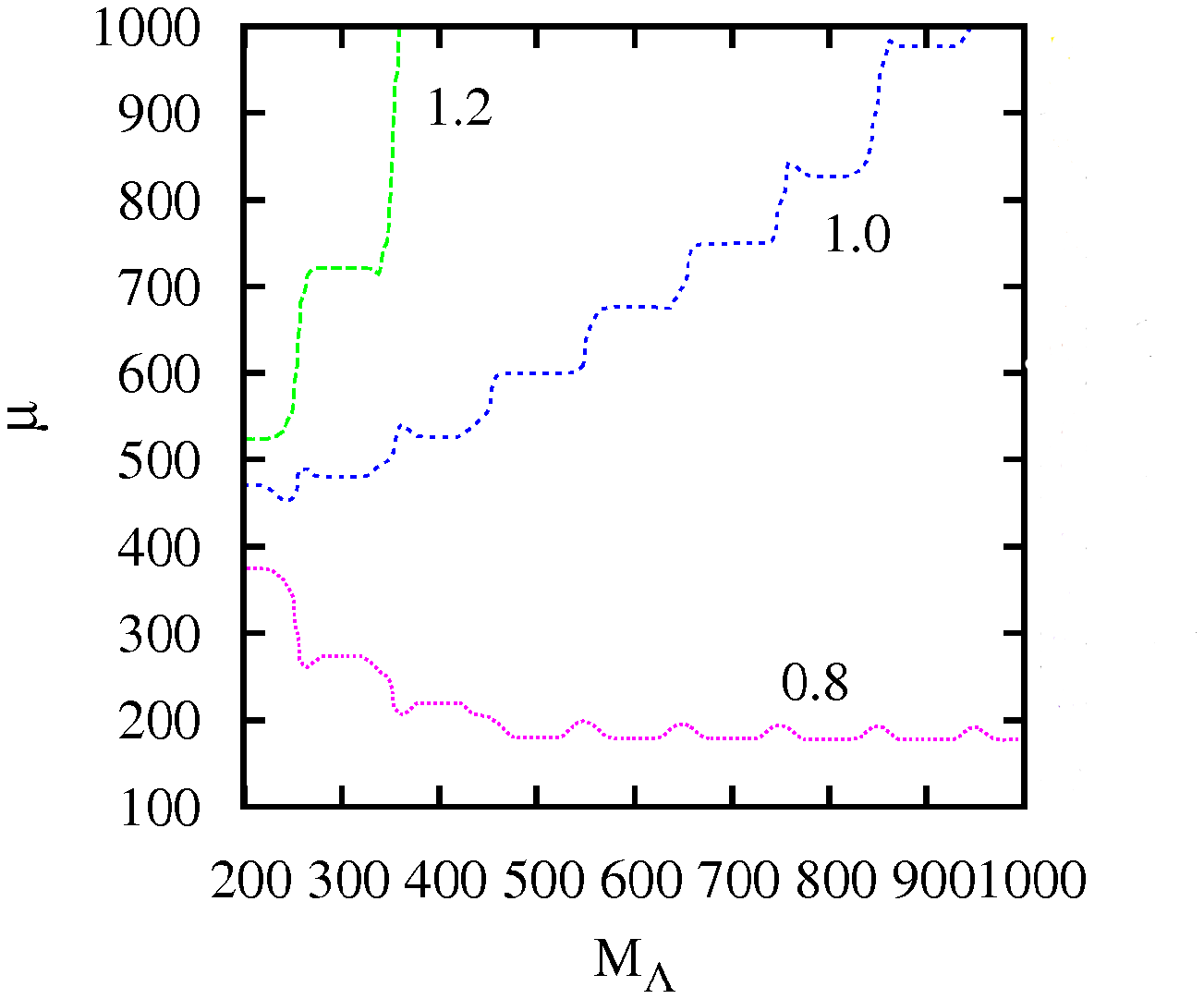}
\caption{ { \footnotesize Contours of constant $R_{\gamma\gamma}$ in (left) $M_2 - \mu$
  and (right) $m_A - \mu$ space. This plot uses the values $M_1 = 50$
  GeV,$\tan \beta = 40$, $A_t = 2.5$ TeV for illustration.  The value
  of $m_A=300$ GeV in the left panel and the value of $M_2 = 200$ GeV
  in the right panel is. The pink, blue and green contours correspond
  to $R_{\gamma\gamma}$ values of 0.8, 1.0 and 1.2
  respectively.}  \label{fig3}}
\end{figure}

For investigating the invisible Higgs branching ratio, we fix our
attention primarily on the $\gamma\gamma$ channel as it currently has
much better statistics than the $4\ell$ channel.  The SM Higgs decay
width into two photons is dominated by the W-boson and top loops.  In
the MSSM case, extra contributions from stop/sbottom and chargino
loops also play a crucial role.  The last mentioned diagram
contributes appreciably only when the lighter chargino is not too
heavy (to avoid mass suppression), and at the same time is a nearly
equal admixture of gaugino and Higgsino states (to maximize the
coupling).  This leaves us with a rather small region with $M_2 \simeq
\mu <$ 200 GeV, where the results are at all sensitive to $M_2$.
Otherwise, the dependence of $R_{\gamma\gamma}$ on $M_2$ is hardly
noticeable.  This is reflected in the left panel of Fig.~\ref{fig3}.
The second panel, on the other hand, confirms that there is
considerable sensitivity of $R_{\gamma\gamma}$ on $m_A$, due to it's
effect on the W-loop contribution via the $hWW$ couplings.  The
dependence on $\mu$ also, is significant, as both the panels show.
One of the reasons is because of the contributions of the sbottom loop
to the $h\longrightarrow \gamma \gamma$ decay width.  As we have set
the parameter $A_b$ to zero, the large values of $\mu$ serve as large
off-diagonal terms in the sbottom mass matrix and result in one
low-mass sbottom state.  The sbottom loop contribution to both Higgs
production and its decay into two photons goes up in the
process. Consequently, one is able to have $R_{\gamma\gamma}$ close to
unity and above, by enhancing the value of $\mu$.

The illustrative Fig.~\ref{fig3} uses $A_t =$ 2.5 TeV, $M_1 = 50$,
$M_2 = 200$ and $\tan\beta =$ 40.  If the SUSY prediction has to be
near about what is predicted by the SM, the region corresponding to 0.8
$< R_{\gamma\gamma} <$ 1.2  mark a ``favoured'' band for the purely
phenomenological MSSM.  Besides this particular case, our scan points
to the following broad conclusions.  For $m_A$ = 300 GeV and $A_t$ =
1.5 TeV, we find that $R_{\gamma\gamma} < 0.8$ irrespective of $\tan
\beta$.  Increasing $A_t$ corresponds to increasing $R_{\gamma\gamma}$
and therefore, we have the appearance of SM-like regions as $A_t$ is
increased to 2.5 TeV.  We also see from the right panel of
Fig.~\ref{fig3} that larger values of $m_A$ increase the value and at
the same time decrease the variation in $R_{\gamma \gamma}$ with $\mu$
and result in regions with stable $R_{\gamma\gamma}$.  Therefore, for
$m_A$ = 1 TeV, (which also corresponds to approaching the decoupling
limit) we have SM-like region even at lower values of $A_t$ = 1.5 TeV.

Given the distributions of $R_{\gamma \gamma}$ for points in the MSSM
parameter space in the allowed Higgs mass range, we now pose the
following questions.  First, if the ATLAS observed signal strengths
are confirmed with more data, is it possible to still have a
significant invisible Higgs branching ratio? And, second, if further
data is more SM-like, what does that mean for the invisible Higgs decay?

\begin{figure}[!htb]
\centering
\includegraphics[width=80mm]{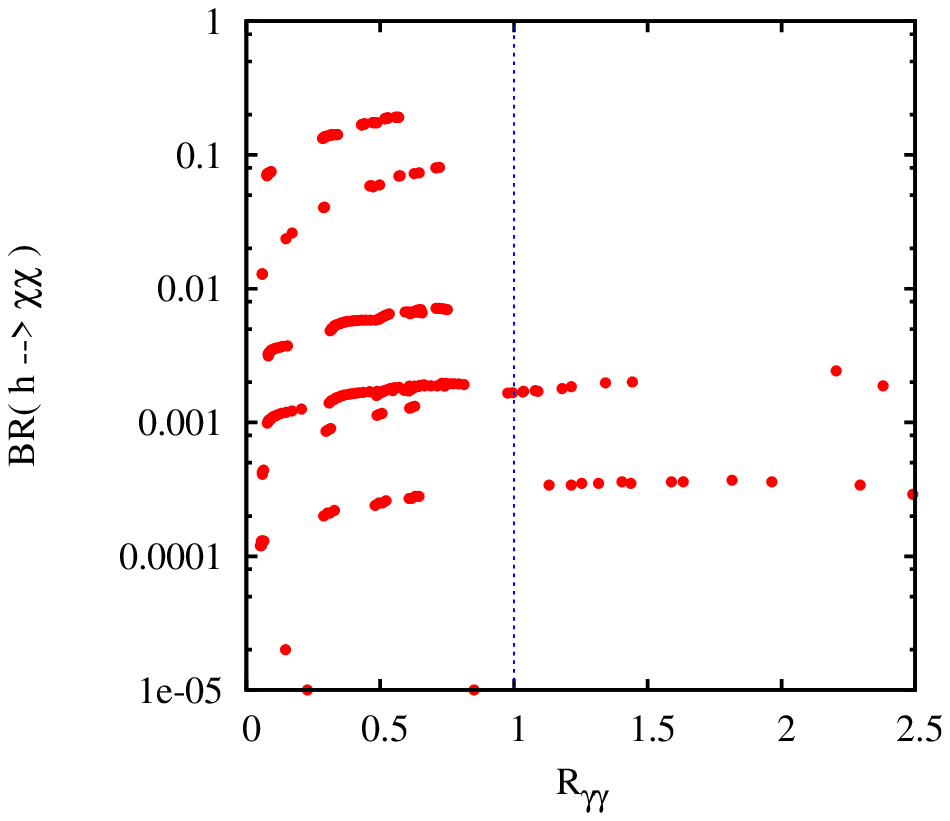}
\includegraphics[width=80mm]{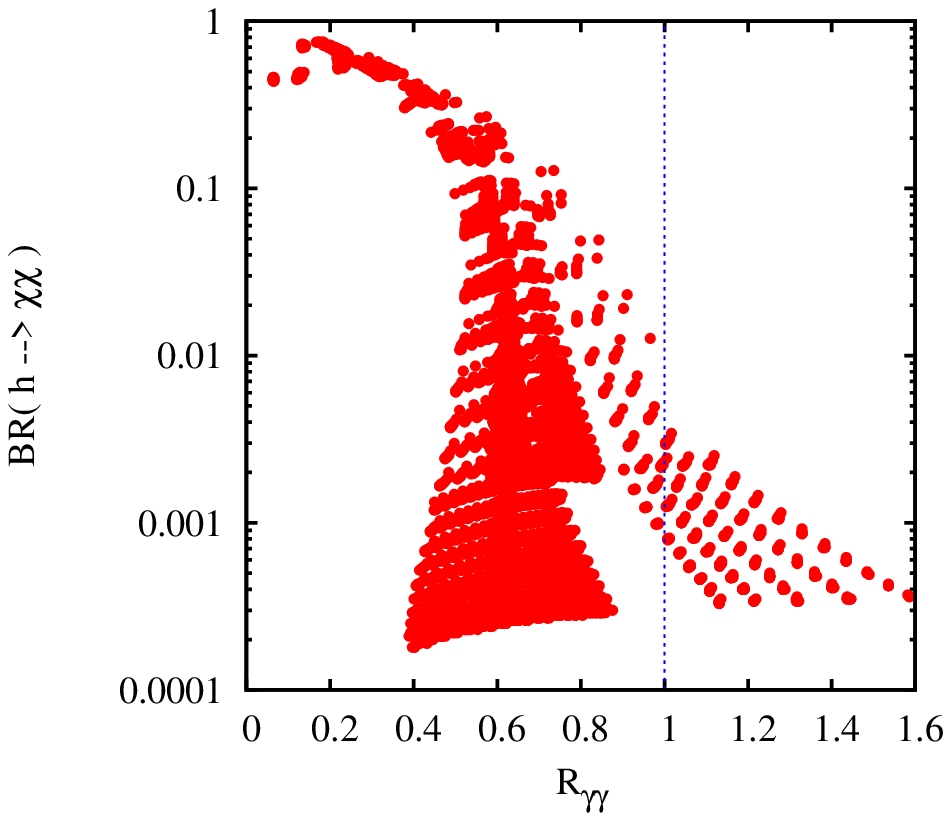}
\includegraphics[width=80mm]{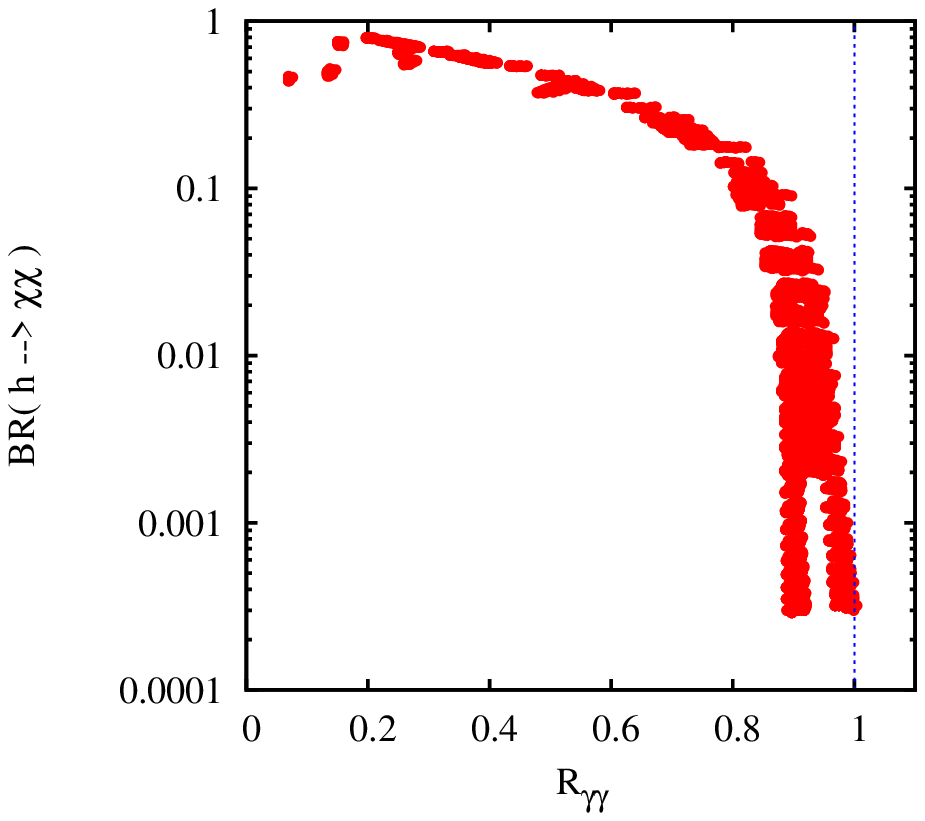}
\caption{{ \footnotesize $R_{\gamma\gamma}$ vs BR($h\longrightarrow \chi^0_1 \chi^0_1$) plots for
 (a) $m_A <$ 300 GeV, (b) $m_A$ = 300 GeV and (c) $m_A$ = 1 TeV. The blue line
 corresponds to SM value of $R_{\gamma\gamma}$ = 1.} \label{fig4}} 
\end{figure}

We show the correlation of $R_{\gamma\gamma}$ and $h\longrightarrow
\chi^0_1 \chi^0_1$ in Fig.~\ref{fig4}.  In general, we notice a
negative correlation between the two quantities.  The first panel of
Fig.~\ref{fig4} corresponds to the region $m_A<300$ GeV which we found
to be able to reproduce current ATLAS data.  However, we find that the
region that agrees best with experiment also corresponds to very small
values of BR($h \longrightarrow \chi^0_1 \chi^0_1$).  Therefore, if
the current results continue to hold with more data, we do not expect
a large invisible Higgs branching fraction.  Large invisible fractions
of the order of 10\% are allowed for $R_{\gamma\gamma}$ in the region
0.6 - 0.8, for all values of $m_A$ as can be seen by comparing the
three panels.  For $m_A = 1$ TeV (third panel), we see that even
$R_{\gamma\gamma}$ upto 0.9 have many points with up to 10\% invisible
branching fractions.

On the whole, the our analysis suggests that {\em it is possible to
  have SUSY contributions to the $\gamma\gamma$ rate at a level
  comparable to that in the SM, and at the same time allow for an
  appreciable invisible decay width, if one is faced with a light
  neutralino dark matter candidate.}  However, if the $\gamma \gamma$
rate is larger than the SM rate, any invisible component is likely to
be very small and undetectable.

\section{Summary and conclusions}

We have performed a completely phenomenological analysis, based on
MSSM without any model assumptions, of the recent data on Higgs search
in the $\gamma\gamma$ and $4\ell$ final states, demanding that the
lightest neutral Higgs mass be in the range 123 - 127 GeV. The very
condition of this mass restriction imposes a considerable constraint
on the parameter space when two-loop corrections to the scalar
potential are included.  Further, we have analysed the parameter space
to identify regions where the $\gamma\gamma$ rate is close to what is
expected with standard model Higgs of mass 125 GeV.  We also compare
the $\gamma\gamma$ and $4\ell$ rates from the allowed MSSM parameter
space to the observed data and to the expected SM rates.  Once the
Higgs mass is decided, the two quantities that most crucially affect
the $\gamma \gamma$ rates are $\mu$, the Higgsino mass parameter, and
$m_A$, the neutral pseudoscalar mass.  It is also found that the role
of loops driven by the lighter sbottom state can be rather
important. In order to see whether there is an irreconcilable tension
between a large $\gamma\gamma$ rate and the invisible decay of the
Higgs into a pair of LSP, we have deliberately confined ourselves to
the case where the lightest neutralino is light. It is found that,
inspite of a mild anti-correlation between the two effects, one can
still have $R_{\gamma\gamma}$ not too far away from unity and at the
same time the invisible Higgs decay branching ratio around 10\% in
certain regions of the MSSM parameter space.

When a positive signal for the Higgs boson is seen, and the rate in a
suppressed channel like the two-photon one is found to be close to
what the standard model predicts, it is common to assume that a new
physics scenario that entails new decay modes of the Higgs is
disfavoured.  Our study reveals that it is not so in the
phenomenological MSSM, and that even measurable invisible decay widths
of the lightest neutral Higgs can coexist with otherwise SM-like
signals. This applies even to the case where R-parity is violated in
SUSY and the LSP is liable to decay.

\section{Acknowledgements}
The authors would like to thank Satyanarayan Mukhopadhyay for useful discussions at the initial stage 
of this work. This work was partially supported by funding available from the Department of Atomic Energy, 
Government of India for the Regional Centre for Accelerator-based Particle Physics, 
Harish-Chandra Research Institute.

\end{document}